\documentclass[reprint,amsfonts,amsmath,amssymb,aps,pra,superscriptaddress,10pt]{revtex4-2}
\usepackage{graphicx}
\usepackage{sansmathfonts}
\usepackage[colorlinks=true,linkcolor=black,anchorcolor=black,citecolor=black,filecolor=black,menucolor=black,runcolor=black,urlcolor=black]{hyperref}

\makeatletter
\renewcommand*{\fnum@figure}{{\normalfont\bfseries \figurename~\thefigure}}
\renewcommand*{\@caption@fignum@sep}{ \textbar{} }
\makeatother
\let\oldcaption\caption
\renewcommand\caption[2][]{\oldcaption[#1]{\textbf{#1.} #2}}

\begin{document}

\title{Large-Scale Tree-Type Photonic Cluster State Generation with Recurrent Quantum Photonic Neural Networks}

\author{Jacob Ewaniuk}
\email{jacob.ewaniuk@queensu.ca}
\affiliation{Centre for Nanophotonics, Department of Physics, Engineering Physics \& Astronomy, 64 Bader Lane, Queen's University, Kingston, Ontario, Canada K7L 3N6}

\author{Bhavin J. Shastri}
\email{bhavin.shastri@queensu.ca}
\affiliation{Centre for Nanophotonics, Department of Physics, Engineering Physics \& Astronomy, 64 Bader Lane, Queen's University, Kingston, Ontario, Canada K7L 3N6}
\affiliation{Vector Institute, Toronto, Ontario, Canada, M5G 1M1}

\author{Nir Rotenberg}
\email{nir.rotenberg@queensu.ca}
\affiliation{Centre for Nanophotonics, Department of Physics, Engineering Physics \& Astronomy, 64 Bader Lane, Queen's University, Kingston, Ontario, Canada K7L 3N6}

\date{\today}

\begin{abstract}
Large, multi-dimensional clusters of entangled photons are foundational resources for scalable quantum technologies, from universal measurement-based quantum computation to global quantum communication networks. Here, we introduce a fundamentally new architecture and protocol for their generation based on recurrent quantum photonic neural networks (QPNNs), with a focus on tree-type cluster states. Unlike existing approaches, QPNN-based generators are not constrained by the the coherence times of quantum emitters or by probabilistic multi-photon operations, enabling arbitrary scaling only limited by loss (which, unavoidably, also affects all other methods). We demonstrate that a single QPNN can learn to perform the full suite of operations required to construct cluster states, from photon routing to entanglement generation, all with near-perfect fidelity and at loss-limited rates, even when built from imperfect components. Our analysis shows that current integrated photonic technologies can already support the generation of 60-photon cluster states, with 100s of photons achievable through modest reductions in component loss. We further evaluate the application of these states in a one-way quantum repeater, identifying performance thresholds for achieving global-scale quantum communication, and highlighting the potential of the QPNN to play a vital role in emerging high-impact quantum technologies.
\end{abstract}

\maketitle

\section{Introduction} \label{sec:intro}
The quantum internet promises unconditionally secure communication \cite{Gisin:07, Gisin:02, Pan:24}, and serves as a foundation for distributed quantum computing \cite{Grover:97, Cirac:99, Main:25} and sensing \cite{Zhao:21, Kim:24}. At its core, such a network must distribute entanglement \cite{Humphreys:18} across distances ranging from kilometers to global scales, a task achievable only by using light. Yet, even in ultra-low-loss telecom fiber, only 300,000 out of one million photons survive a 30~km journey, and just 8 in a million reach a node 300~km away. These severe transmission losses are compounded by the no-cloning theorem, which prohibits amplification of quantum signals \cite{Wootters:82}, indicating that scalable quantum networks cannot be realized simply by increasing the photon flux.

Instead, one can imagine encoding individual quantum bits (qubits) of information into entangled multi-photon states, known as cluster states \cite{Raussendorf:03}, with built-in redundancies that enable indirect measurement of lost photons through those that remain. Among these, tree-type cluster states---two-dimensional graphs of maximally entangled photons \cite{Varnava:06}---have been identified as promising resources for both one-way \cite{Borregaard:20, Wo:23} and two-way \cite{Hilaire:21} quantum repeaters, and may also enable loss-tolerant, measurement-based quantum computation \cite{Varnava:07, Briegel:09}. Consequently, a range of protocols and recent experiments have been developed to generate such states using linear quantum optics \cite{Browne:05, Bodiya:06} or quantum emitters such as quantum dots, defect centers or atoms \cite{Buterakos:17, Zhan:20, Thomas:24}. Each approach leverages the strengths of its platform, from the ability to integrate linear circuits on photonic chips \cite{Carolan:15, Bao:23}, to deterministic processing using quantum emitters \cite{Cogan:23, Uppu:21, Huet:25}. Nevertheless, as we discuss below, all current protocols suffer from fundamental limitations that hinder the scalable generation of large photonic cluster states.

\begin{figure*}[ht!]
\centering
\includegraphics{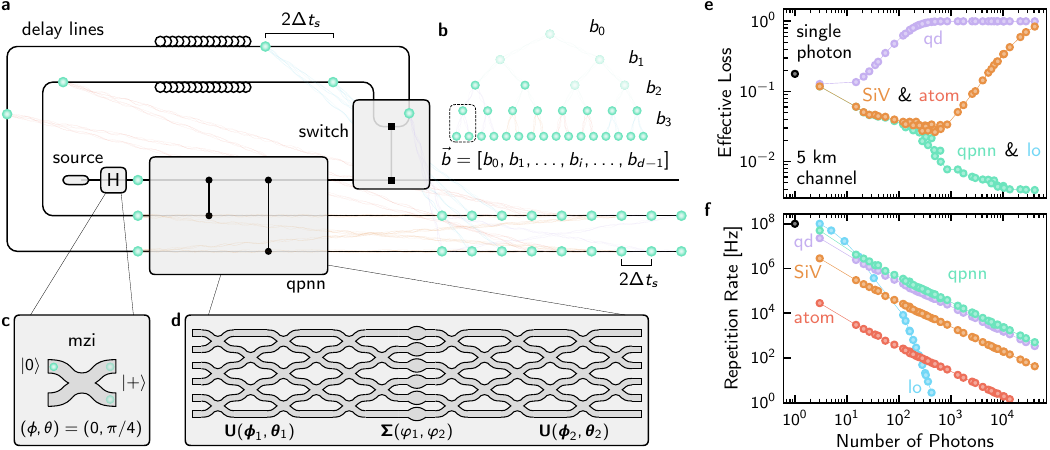}
\caption[Tree-type cluster state generator based on a recurrent quantum photonic neural network (QPNN)]{\textbf{a} Architecture of the tree state generator. One single-photon source (that can emit each $\Delta t_s$) and Hadamard gate (H) prepare and inject each photon of the tree into the generator, where they are subsequently routed through the QPNN, switch, and delay lines according to a timing protocol for the desired tree shape. Each wire corresponds to a single photonic qubit, thus two spatial modes in the dual-rail encoding scheme. \textbf{b} Example tree-type photonic cluster state, with branching vector $\vec{b} = \left[2, 2, 2, 2\right]$, generated in \textbf{a}. Faded photons are yet to be generated in \textbf{a}, while those present can be identified by the colors of the lines connecting them, all of which represent maximal entanglement. One unit cell of the tree is outlined by a dashed box. \textbf{c} Inset of the Hadamard gate in \text{a}, realized by a Mach-Zehnder interferometer (MZI) with two phase shifters $(\phi, \theta)$, where each emitted photon in state $\left|0\right\rangle$ is transformed to state $\left|+\right\rangle$ by selecting phase shifts $(\phi, \theta) = (0, \pi / 4)$. \textbf{d} 2-layer, 6-mode QPNN that operates on three photonic qubits as shown in \textbf{a}. Each linear layer, $\mathbf{U}(\boldsymbol{\phi}_i, \boldsymbol{\theta}_i)$, is formed by a mesh of MZIs and separated by single-site nonlinearities $\boldsymbol{\Sigma}(\varphi_1, \varphi_2)$. \textbf{e}, \textbf{f} Comparison of tree generation protocols based on QPNNs (this work), active control of different quantum emitters (qd, SiV, atom), and linear optics (lo), assuming that each photon in each tree has a $10\%$ chance of being lost in the generator, regardless of the tree size (i.e. number of photons), then traverses a 5 km fiber channel, accruing $\sim 18\%$ additional loss. In \textbf{e}, \textbf{f} respectively, the effective loss of the logical qubit and repetition rate of the generator are shown as the tree scales. Further details on this comparison are given in Sec.~S1 of the Supplementary Information.}
\label{fig:generator}
\end{figure*}

Here, we introduce a novel generator for tree-type photonic cluster states (Fig.~\ref{fig:generator}a),
 based on a recurrent quantum photonic neural network (QPNN). QPNNs are reconfigurable, nonlinear quantum photonic circuits that leverage machine learning to learn deterministic quantum state transformations \cite{Steinbrecher:19}, even in the presence of experimental imperfections \cite{Ewaniuk:23}. By combining a QPNN with a single-photon source, optical switches, and delay lines, the system can recursively entangle emitted photons to construct arbitrary tree structures (Figs.~\ref{fig:generator}a–b). To contextualize the advantages of our approach, we benchmark its performance against representative linear-optical \cite{Bodiya:06} and emitter-based \cite{Zhan:20} protocols (Figs.~\ref{fig:generator}e–f), while assuming ideal photon sources and a uniform 10\% per-photon loss rate during generation (independent of tree size; see Supplementary Sec.~S1), followed by a 5~km channel. We find that both linear-optical protocols (blue) and our QPNN protocol (green) exhibit monotonically decreasing logical qubit loss with increasing cluster state size (Fig.~\ref{fig:generator}e). In contrast, the size of cluster states generated using emitter-based protocols are limited by decoherence, which acts to increase the effective loss at about 15 photons when using quantum dots (purple), with a similar performance degradation for atoms and defect centers (red and orange, respectively) at about 1000 photons. In terms of repetition rate, the QPNN and emitter-based protocols scale as $1/n$ with the number of photons $n$, while linear-optical approaches suffer exponential slowdown (Fig.~\ref{fig:generator}f). The QPNN thus offers deterministic performance and scalability unconstrained by the coherence-time limitations of emitter-based protocols.

In what follows, we present the architecture of the proposed generator, introduce a timing protocol adaptable to arbitrary tree geometries, and describe a training procedure for the QPNN, demonstrating that fidelities approaching unity are achievable with current technology. Using a trained network model, we evaluate system performance under scaling and realistic imperfections. Finally, we apply our generator to a one-way quantum repeater protocol and demonstrate its potential to significantly enhance communication rates, particularly as component technologies improve. Collectively, these results suggest that QPNN-based generators could enable the efficient formation of large-scale photonic cluster states, offering a promising path toward scalable quantum networks.


\section{Generation Protocol} \label{sec:protocol}
\begin{figure*}[ht!]
\centering
\includegraphics{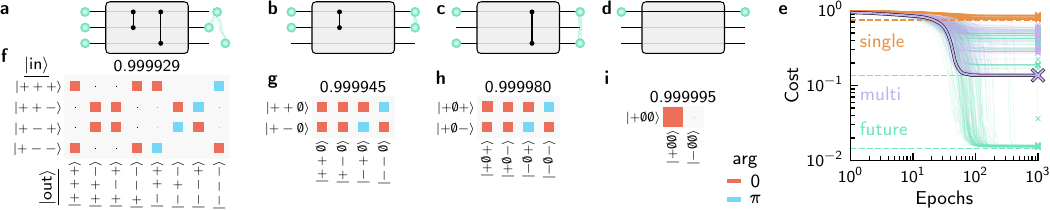}
\caption[Training a 2-layer, 6-mode QPNN as a tree-type photonic cluster state generator]{\textbf{a}-\textbf{d} Circuit diagrams for each QPNN operation required by the generation protocol. \textbf{e} Minimization of the network cost (i.e. average error) during 200 optimization trials of 1000 epochs each for the \textit{single}-, \textit{multi}- and \textit{future}-platform (see main text and Sec.~\ref{sec:methods} for platform details). Dashed lines denote the loss limit (i.e. minimum achievable cost due to loss). \textbf{f}-\textbf{i} Hinton diagrams resolved in the X-basis for each operation of the \textit{multi}-platform QPNN outlined in black in \textbf{e}, where the uppermost photonic qubit is $\left|+\right\rangle$ at the input (vertical axis), yet can belong to a superposition of $\left|+\right\rangle$ and $\left|-\right\rangle$ at the output (horizontal axis). Each box is colored according to its argument, which is always within $\pi/100$ of either $0$ or $\pi$ up to an insignificant global phase. When a photonic qubit is missing at any input or output port of the network, $\emptyset$ is written in its place. The fidelity of each operation is given above its Hinton diagram, never falling below 0.999929 for this network.}
\label{fig:training}
\end{figure*}

The key to our approach is that the QPNN generates unit cells of a tree-type photonic cluster state (e.g., the three photons in the dashed region of Fig.~\ref{fig:generator}b, where $b_3 = 2$ specifies two branches), and then recursively applies this operation to construct the full tree. In each unit cell, vertices represent a photon initialized in the $\left|+\right\rangle$ state, while edges denote entanglement formed by applying a controlled-Z gate between connected photons \cite{Raussendorf:03}. In the illustrated case, the QPNN must entangle 3 photons at a time, although as we show in Supplementary Fig.~S1, this operation can be generalized to accommodate arbitrary branching ratios. A representative circuit for a uniform branching factor of $b = 2$ (i.e., $b_j = 2$ for all $b_j$ in $\vec{b}$) is shown schematically in Fig.~\ref{fig:training}a.
To ensure resilience to photon loss within the device, the QPNN must also be capable of entangling fewer photons (for instance, two photons as shown in Figs.~\ref{fig:training}b,c), to ensure the integrity of the tree \cite{Varnava:06}. In the initial generation steps, when only a single photon may be present (as defined by the timing protocol described below), the network must function as an identity operator, simply transmitting the photons unchanged (Fig.~\ref{fig:training}d). The QPNN must therefore select the appropriate transformation based on the incoming configuration of photons.

To train the QPNN, we first construct each linear interferometric mesh (cf. $\mathbf{U}(\boldsymbol{\phi}_i, \boldsymbol{\theta}_i)$ in Fig.~\ref{fig:generator}d) using imperfect components, following the methodology described in Ref.~\cite{Ewaniuk:23}. This approach incorporates realistic nonidealities, including unbalanced losses and imperfect routing, as typically encountered in integrated photonic circuits (see Methods Sec.~\ref{sec:methods} for further details on network construction and training). Here, we construct three QPNN models, all of which have $(50 \pm 0.5)\%$ directional couplers, but differing in Mach-Zehnder interferometer (MZI) loss: the \textit{single} network is constructed using the best-performing, monolithic integrated photonic platform available, yielding $0.2130 \pm 0.0124$ dB loss per MZI \cite{Alexander:25}; the \textit{multi} network is constructed from state-of-the-art, individually demonstrated photonic elements, resulting in a loss of $0.0210 \pm 0.0016$ dB per MZI \cite{Alexander:25, Parra:20, Harris:14}; the \textit{future} network assumes an order-of-magnitude reduction in loss from the current best (i.e., $0.00210 \pm 0.00016$ dB per MZI), which allows us to explore QPNN operation as photonic fabrication improves. We further assume a Kerr-type nonlinearity that imparts a photon-number dependent phase shift without introducing wavepacket distortions. While yet to be realized experimentally, this nonlinear transformation can be implemented using cavity-assisted light-matter interactions involving incident photons and a $\Lambda$-type three-level emitter, using a protocol that may be implemented in integrated photonic platforms \cite{Basani:24}.

We train 200 QPNNs for each network model over 1000 training epochs and present the results in Fig.~\ref{fig:training}e. Across all models, the trained QPNNs can achieve loss-limited operation, indicated by the dashed lines. Specifically, the \textit{single} model achieves a final cost within 0.0115 of the loss limit (see Methods Sec.~\ref{sec:methods}), while the \textit{multi} and \textit{future} models reach within 0.0014 and 0.0004 respectively. These results indicate that the QPNNs successfully learn to overcome imperfections and achieve balanced loss operation. Notably, the \textit{multi} model (purple) achieves a cost of 0.136 at best when trained to perform all required operations (Figs.~\ref{fig:training}a-d) simultaneously, demonstrating the potential of our devices.

The results are even more encouraging when we consider the expected fidelity of the different operations, shown in Figs.~\ref{fig:training}f-i for the \textit{multi}-model QPNN traced with a dark line in Fig.~\ref{fig:training}e. These Hinton diagrams illustrate the output states (horizontal axis) generated by the network for each input state (vertical axis), with the square sizes encoding amplitudes and color denoting relative phase. Comparisons to the ideal diagrams in the same basis (see Supplementary Fig.~S2, alongside trained diagrams in the Z-basis) confirm the near-perfect operation of the QPNN. The average fidelity across all operations is 99.995\%, indicating that nearly all residual errors arise from photon loss, manifesting primarily as reduced operational throughput. The lowest individual fidelity, $99.993\%$, is observed for the three-photon operation (Fig.~\ref{fig:training}f). Collectively, these results demonstrate that, unlike linear circuits which are designed to perform a single operation, a QPNN can learn to perform multiple, photon-number-dependent operations within a single architecture. As we next show, this versatility opens new routes to complex quantum information processing protocols, including the scalable generation of two-dimensional photonic cluster states.

As an illustrative example, we consider the generation of a depth-2 ($d=2$) tree with a branching factor $b = 2$, resulting in a seven-photon cluster state (see Supplementary Sec.~S3 for a generalization to arbitrarily-shaped trees). The timing protocol requires just three delay lines (cf. Fig.~\ref{fig:generator}a) with delays of $4\Delta t_s$, $3\Delta t_s$, and $2\Delta t_s$ respectively, to ensure that all three photons in each unit cell arrive at the QPNN simultaneously. Here, $1/\Delta t_s$ is the fundamental clock rate of the cluster state generator, which in our implementation is limited by the requisite nonlinearity. Specifically, the cavity-assisted nonlinearity  is predicted to achieve near-unity fidelity for photons with temporal widths on the order of 1~ns \cite{Basani:24}, leading us to set $\Delta t_s = 10$~ns to avoid overlap. This results in delay line lengths of 8.2~m, 6.2~m and 4.1~m, with corresponding losses of $0.0014$~dB, $0.0010$~dB and $0.0007$~dB, respectively (see Methods Sec.~\ref{sec:methods} for details). The required delay length grows exponentially with $d$, resulting in the scaling shown in Fig.~\ref{fig:generator}f. We also require $1\times4$ and $1\times2$ optical switches operating at speeds exceeding $1/\Delta t_s = 100$~MHz, which are readily achievable using electro-optic switches that can reach bandwidths above 10~GHz and insertion losses of only 0.107~dB per switching stage \cite{Alexander:25}.

\begin{figure}[ht!]
\centering
\includegraphics{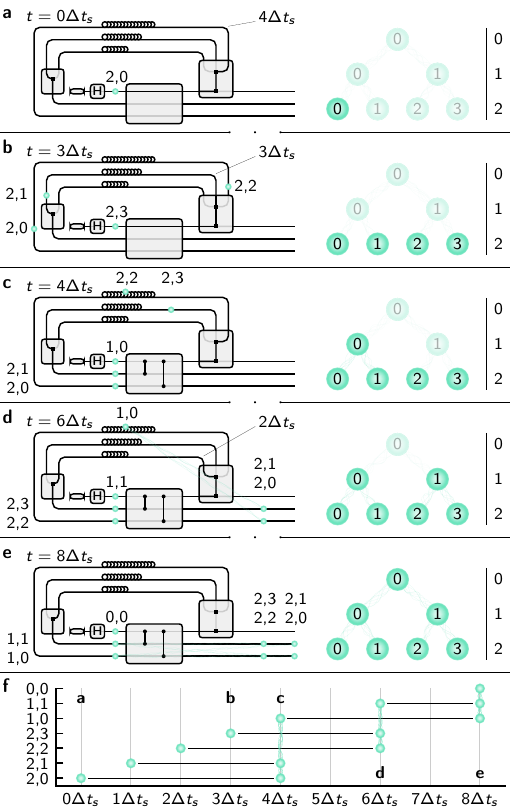}
\caption[Timing protocol for tree-type photonic cluster state generation]{Procedure for generating a tree state with branching vector $\vec{b} = \left[2, 2\right]$. \textbf{a}-\textbf{b} Individual photons from row 2 of the tree are emitted in subsequent timesteps $t$, separated by $\Delta t_s$, the time between source triggers, then traverse through the QPNN which acts as an identity operation. The switch is adjusted at each timestep, directing each pair of consecutive photons to each of the two longest delay lines ($4\Delta t_s$ and $3\Delta t_s$), such that the latter photon catches up to the former. \textbf{c}-\textbf{d} Delayed photons arrive at the QPNN input with a newly emitted photon such that all three are subsequently entangled. The top photon is routed to a delay line by the switch while the other photons reach the output of the generator. These operations are separated by 2$\Delta t_s$, and the shortest delay line ($2\Delta t_s$) is selected at $t=6\Delta t_s$ to accommodate this change. \textbf{g} The root photon of the tree, $(0, 0)$, is emitted such that it arrives at the QPNN simultaneously with the delayed photons. The switch ensures the root is also routed to the output after it becomes entangled with the others. \textbf{f} Each dark line specifies the timesteps that a photon, as labeled on the vertical axis, is traversing the generator. Photon markers denote timesteps where the QPNN acts on a given photon, with connecting lines added to indicate entangling operations.}
\label{fig:protocol}
\end{figure}

Representative steps for the cluster state generation protocol are shown in Fig.~\ref{fig:protocol}.
The tree is constructed from the bottom up, with the first four timesteps dedicated to the emission of the bottom row of photons. During these steps, the QPNN acts as an identity operator, while the switch alternates between the $4\Delta t_s$ and $3\Delta t_s$ delay lines. For example, the first photon $\left(2,0\right)$ is routed into the longer delay line at $t = 0$ (Fig.~\ref{fig:protocol}a), whereas the fourth photon $\left(2,3\right)$ is routed into the shorter delay at $3\Delta t_s$ (Fig.~\ref{fig:protocol}b). At the next timestep, $t=4\Delta t_s$ (Fig.~\ref{fig:protocol}c), the first photon of the middle row $\left(1,0\right)$ is emitted and enters the QPNN simultaneously with the first two (delayed bottom) photons. Receiving three inputs, the QPNN performs a three-photon entangling operation (cf. Figs.~\ref{fig:training}a,f), forming the first complete unit cell. The newly generated head photon is routed into the $4\Delta t_s$ delay, while the two entangled photons are released. Two timesteps later, at $t=6\Delta t_s$ (Fig.~\ref{fig:protocol}d), the second middle-row photon $\left(1,1\right)$ is emitted, entangled with the remaining bottom-row photons, and routed into the $2\Delta t_s$ delay line. Finally, the root photon $\left(0,0\right)$ is emitted at $t=8\Delta t_s$ (Fig.~\ref{fig:protocol}e), arriving at the QPNN together with the two delayed middle-row photons, completing the entanglement of the seven-photon tree. Altogether, the full 7-photon cluster state is generated within $8\Delta t_s =80$~ns. Notably, this duration can be further reduced by incorporating additional photon sources or delay lines.

\begin{figure}[ht!]
\centering
\includegraphics{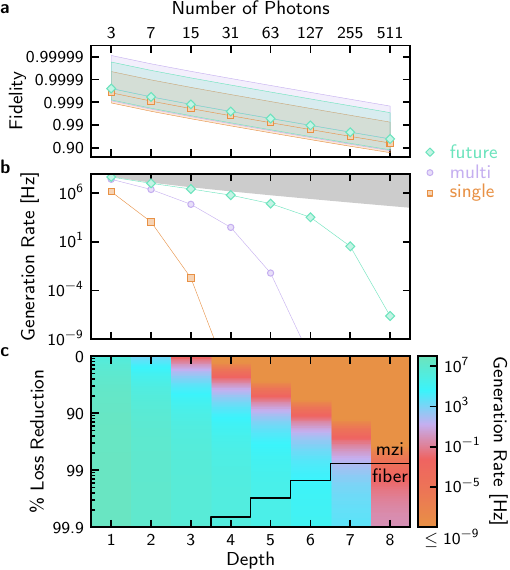}
\caption[Tree-type photonic cluster state generator performance for 2-branch-only trees of increasing depth]{\textbf{a} Fidelity of tree states of increasing depth, with the mean (95\% confidence intervals) of fit beta distributions shown by markers (shaded regions), when generated using a \textit{single}-, \textit{multi}- and \textit{future}-platform QPNN, respectively (cf. Fig.~\ref{fig:training}). \textbf{b} The corresponding rate at which entire trees (i.e. not missing any photons) can be generated for the three platforms as a function of the tree depth, assuming that the QPNN is trained to loss-limited operation. The shaded region is unattainable, even with lossless components, due to the time needed to generate all of the individual photons in the protocol. \textbf{c} Generation rate for different depth trees as a function of the percent loss reduction starting from the \textit{single}-platform QPNN (i.e. at 0 reduction), allowing us to understand the network performance if all elements are improved together. The solid black line separates the regime where MZI losses dominate from the one where fiber losses dominate.}
\label{fig:growing}
\end{figure}

Next, we apply our protocol to construct cluster states of increasing size by scaling the tree depth $d$ while maintaining a branching factor $b=2$. Using the QPNNs trained in Fig.~\ref{fig:training}, we simulate the generation of progressively larger trees and summarize the results in Fig.~\ref{fig:growing}. These calculations incorporate not only fiber losses and QPNN-related errors, but also system-level imperfections, including losses from couplers and switches. The complete list of imperfections included for each model is provided in Supplementary Table~S1.
In Fig.~\ref{fig:growing}a, we observe that the average state fidelity remains $> 90\%$ across all three QPNN models, \textit{single}, \textit{multi}, and \textit{future}, even for cluster states containing up to 511~photons. Remarkably, the fidelities of the three QPNN models are nearly indistinguishable, indicating that MZI loss, despite being higher in the \textit{single} model, is no longer the dominant error contribution. Instead, fidelity is now limited primarily by photon routing errors. This is further supported by Supplementary Fig.~S6, where fidelity is shown to improve with increasingly balanced directional couplers. To reach a $99\%$ fidelity threshold for a 511-photon cluster ($d=8$), the routing errors must remain below 0.25\%, which is approximately half the error level achieved by current state-of-the-art devices \cite{Alexander:25}.

Although high-fidelity cluster states can be generated at large scales, the corresponding generation rate decreases rapidly with tree size. As shown in Fig.~\ref{fig:growing}b, the rate of generating a complete tree state, i.e., one with no photons lost, declines across all three QPNN models as the tree grows. This decrease is driven by two primary factors. First, generating more photons simply requires more time, leading to an inherent $1/n$ scaling of the rate with photon number (cf. shaded region in Fig.~\ref{fig:growing}b). Second, the cumulative effect of photon loss increases with scale, contributing to the observed model-dependent differences in generation rate.

There are strong reasons for optimism. Even with the current \textit{single}-platform model (orange line in Fig.~\ref{fig:growing}b), our QPNN-based approach is expected to generate 7-photon trees ($d = 2$) at a rate of 1~kHz (including all loss), equivalent to the rate expected using the atom-based protocol of Ref.~\cite{Thomas:24} with a perfect setup, yet five orders-of-magnitude higher than the experimental demonstration. Note that this particular atom-based protocol relies on probabilistic operations, meaning that our approach would scale favorably in comparison. As low-loss components become further integrated (in the \textit{multi}-platform model, purple), tree sizes can more than double to 15 photons, or even increase by nearly an order-of-magnitude to 63 photons, with corresponding generation rates of 73~kHz and 6~mHz, respectively. A further order-of-magnitude reduction in loss (\textit{future} platform, green) enables dramatic improvements, boosting the generation rate of 63-photon trees by nearly 7 orders-of-magnitude to roughly 86~kHz, and allowing 255-photon trees to be generated at nearly 5~Hz. These findings highlight how even modest reductions in photonic losses can yield substantial improvements in device performance and establish concrete, realistic requirements for photonic component fabrication.

An important complementary question is how much the current best integrated platform (\textit{single}) must be improved to enable rapid generation of cluster states at larger scales. In Fig.~\ref{fig:growing}c, we address this by plotting the generation rate as a function of the fractional reduction in losses from the \textit{single} platform baseline \cite{Alexander:25}. In this analysis, the losses from MZIs, switches and couplers are uniformly scaled by this factor, while fiber losses remain fixed. The results show that with a 99\% reduction in on-chip losses, comparable to the difference between the \textit{single} and \textit{future} MZI values, cluster states with over 100 photons could be generated at 20~kHz, and states approaching 500 photons would be reachable. As indicated by the black contour line in Fig.~\ref{fig:growing}c, generation rates for larger states are limited primarily by fiber loss, whereas MZI-related losses dominate for smaller states. As discussed below, this limitation may be alleviated through alternative architectures, such as those employing several photon sources.


\section{Boosting Communication Rates} \label{sec:rates}
\begin{figure}[ht!]
\centering
\includegraphics[width=\columnwidth]{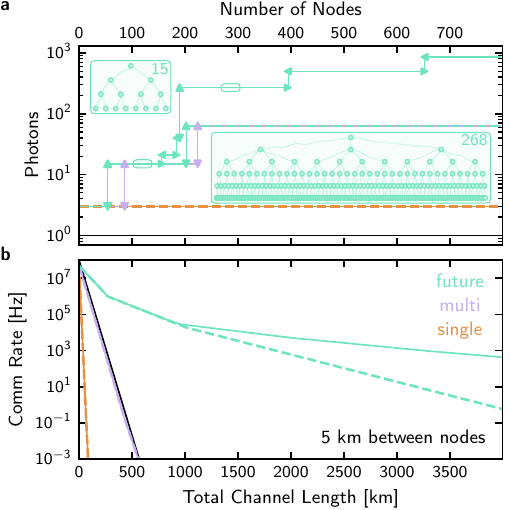}
\caption[Projections for a one-way quantum repeater based on a QPNN-based tree state generator]{\textbf{a} The number of photons in the optimal tree-type cluster state as a function of the total channel length (keeping a constant 5 km node separation; number of nodes shown on top axis) for the three different QPNN platforms (cf. Fig.~\ref{fig:training}). Dashed curves represent trees with a constant branching ratio $b=2$, while solid curves are arbitrarily-shaped trees with $\max\{\vec{b}\}\leq 4$. A change in the depth (branching) of the tree shape is denoted by vertical (horizontal) arrowheads. Two exemplary trees are shown in insets for the future QPNN, comprised of 15 and 268 photons, demonstrating that tree growth is primarily accomplished through increased branching. \textbf{b} The corresponding communication rates for the different clusters in \textbf{a}, benchmarked to the rate using single photons (black line). In this model, we assume that information is transferred between logical qubits perfectly at each node, such that all rate reduction is due to losses. Further details can be found in Supplementary Information Sec.~S5.}
\label{fig:boosting}
\end{figure}

Tree-type cluster states are particularly well-suited for quantum communication applications, prompting the question of whether the states generated in our framework---given their fidelities, generation rates, and sizes---can be practically deployed in this context. To evaluate this, we model the performance of a one-way quantum repeater \cite{Borregaard:20} that employs these tree states as a resource, using the three loss models discussed. In all cases, we assume that repeater nodes are spaced 5~km apart. For each total channel length, we compute the optimal tree state (i.e., the one that yields the fastest communication rate) that can be generated by our QPNNs. The results are shown in Fig.~\ref{fig:boosting}a, both for the constrained case where the maximum branching factor is fixed at $\max\{\vec{b}\}=2$ (dashed curves), as considered in the previous section, and for the general case where the branching may increase arbitrarily (solid curves).
In essence, optimizing performance requires balancing two competing effects: larger cluster states have slower repetition rates and endure more loss during generation, but offer greater tolerance to overall photon loss, thereby increasing the likelihood of successfully transmitting quantum information between nodes. For the \textit{single}-platform QPNN, high loss rates necessitate the use of the smallest possible tree state regardless of channel length ($d=1$ and $b=2$, resulting in 3 photons). In contrast, for both the \textit{multi}- and \textit{future}-platform QPNNs, the optimal tree state size increases with the total channel length. This trend is particularly pronounced for the unconstrained \textit{future}-platform (solid green curve), where the optimal tree state grows to 1000s of photons as the channel length approaches 1000s of kilometers---far exceeding the coherence-limited size of emitter-based protocols. Interestingly, these large states are constructed by increasing the branching factors (i.e., elements of $\vec{b}$) rather than the tree depth $d$, which remains constrained by fiber loss, consistent with the trends observed in Fig.~\ref{fig:growing}c.

In Fig.~\ref{fig:boosting}b, we present the communication rates enabled by the generated cluster states, computed from the corresponding repetition rates and effective losses between nodes (as detailed in Supplementary Fig.~S7), and benchmark them against the communication rate achieved by directly transmitting individual photons. As expected, the small size of the \textit{single}-platform states renders them insufficient for all but the shortest channels, offering no improvement over direct transmission. In contrast, the \textit{multi}-platform states perform comparably to individual photon transmission, supporting channel lengths of up to 235~km at kHz rates. Strikingly, the \textit{future}-platform cluster states can grow to sufficient sizes to enable meaningful communication rates for global-length channels. 

For example, at a channel length of 1000~km, both the constrained and unconstrained cluster states (comprising 15 and 268 photons, respectively) enable communication rates of approximately 20~kHz. Even more remarkably, for a 3000~km channel, while the communication rate of the constrained cluster (now at 63 photons) drops to 20~Hz, the unconstrained tree (comprising 496 photons) continues to support rates above 1~kHz. These results demonstrate that an order-of-magnitude improvement in photonic component losses relative to current platforms would suffice to enable the generation of massively entangled cluster states, paving the way for practical implementations of a global quantum network.


\section{Discussion} \label{sec:discussion}
Looking to the future, it is clear that much remains to be done before a QPNN such as we envision is built and generates massively entangled photonic states. Three key areas demand attention. First, all the separate photonic components must be integrated on a single platform, likely including quantum emitters both for the sources and nonlinearities. Here, improvements are ongoing, both to the underlying photonic platforms \cite{Alexander:25, Bao:23, Adcock:21} and towards hybrid integration \cite{Kim:17, Wan:20, Chanana:22, Larocque:24}. Second, the losses of individual components need to be further reduced if kHz (or even MHz) generation rates of states with more than 30 photons are desired (cf. Fig.~\ref{fig:growing}b). Finally, an efficient nonlinearity on an integrated platform must be demonstrated. This will likely require quantum emitters, for example quantum dots \cite{Staunstrup:24, LeJeannic:22, Nielsen:24} or defects in diamond \cite{Pasini:24, Bhaskar:17, Orphal-Kobin:23, Bhaskar:20} or silicon \cite{Bergeron:20, Prabhu:23}, and protocols such as the cavity-assisted nonlinearity we consider here \cite{Basani:24} or others based on multiple chirally-coupled emitters \cite{Levy-Yeyati:25, Schrinski:22}. Encouragingly, as shown by this work, QPNNs learn to overcome many imperfections, meaning that near-ideal operation is possible even with sub-optimal nonlinearities or photon routing.

There are ways of increasing the performance of the recurrent QPNN beyond simply decreasing losses, albeit at the cost of increased system complexity. More obviously, access to additional fiber delay line lengths would enable one to eliminate empty timesteps from the protocol, where in our examples all photons are in the delay lines (so, for example, at $t=5\Delta t_s$ in Fig.~\ref{fig:protocol}). Here, the increased complexity would be the need for larger switches. Along this vein, the use of multiple sources would significantly increase the operational rate. For example, when $b=2$, the use of 3 sources would allow one to start by generating the entire bottom row of unit cells in the tree, rather than the individual bottom photons. Because the bottom row contains more than half of the photons of the tree, this straightforward change would double generation rates (see Supplementary Fig.~S8). Again, this requires additional switches as well as more emitters. Finally, one could imagine replacing the delay lines with long-lived quantum memories \cite{Mouradian:15, Zhang:23, Bersin:24, Liu:25}, which would both enable a more efficient protocol with no dead time and preclude the need to send photons off-chip during generation.

While we focus on tree states in this work, our QPNN-based approach can likely be extended to other cluster state types with only modest alterations. Specifically, if the cluster can be split into unit cells, we can teach a QPNN to form them, then recursively apply it to build the full state. This procedure is particularly well-suited for the two- and three-dimensional lattices of entangled photons often considered in measurement-based quantum computation \cite{Raussendorf:07a, Raussendorf:07b}. Alternatively, one could envision the unit cells as logical qubits encoded for error correction, that are then entangled with others for processing \cite{Bartolucci:23}. In this vein, more general repeater graph states \cite{Azuma:15} could be formed by multiplexing the tree-generating QPNNs considered here, potentially further enhancing communication rates in quantum networks.

Overall, we present a scheme for large-scale cluster state generation that does not suffer from many of the fundamental limitations of current linear-optical or emitter-based schemes. By using recurrent QPNNs, we propose a system that can learn to overcome many of its own imperfections, and discuss how such a system could enable truly long-distance quantum communications, although we note that the generated states are equally important for other quantum technologies such as quantum computation \cite{Raussendorf:03, Briegel:09}.


\section{Methods} \label{sec:methods}
\noindent\textbf{Optimizing Network Performance}\\
The QPNN performs both linear and nonlinear processing of the photonic qubits, as displayed in Fig.~\ref{fig:generator}d. Each QPNN comprises $m$ modes and $L$ linear layers, each realized by an $m\times m$ mesh of MZIs, where the $i$\textsuperscript{th} layer is programmed by $\boldsymbol{\phi}_i$, $\boldsymbol{\theta}_i$, the phase shifts applied in each MZI of the mesh, to enact the linear unitary transformation $\mathbf{U}(\boldsymbol{\phi}_i, \boldsymbol{\theta}_i)$ \cite{Steinbrecher:19}. Between each pair of consecutive layers, single-site few-photon nonlinearities ($\boldsymbol{\Sigma}(\varphi_1, \varphi_2)$) are applied in each mode, where $n$ incident photons pick up a photon-number-dependent phase shift of $\varphi_1 + (n-1)\varphi_2$ \cite{Basani:24}. Here, we set $\varphi_1 = 0$, $\varphi_2 = \pi$ throughout. By multiplying the constituent linear ($\mathbf{U}$) and nonlinear ($\boldsymbol{\Sigma}$) transformations in order (cf. Fig~\ref{fig:generator}d), we arrive at the system function,
\begin{equation} \label{eq:sys}
   \mathbf{S} = \mathbf{U}(\boldsymbol{\phi}_L, \boldsymbol{\theta}_L)\prod_{i=1}^{L-1}\boldsymbol{\Sigma}(\varphi_1, \varphi_2)\mathbf{U}(\boldsymbol{\phi}_i, \boldsymbol{\theta}_i),
\end{equation}
of an $L$-layer QPNN. Once constructed, the system function can be applied to any given input state $\left|\mathrm{in}\right\rangle_k$ to produce an output state $\left|\mathrm{out}\right\rangle_k = \mathbf{S}\left|\mathrm{in}\right\rangle_k$. To teach a QPNN to perform some desired operation, we train its variational parameters $\boldsymbol{\phi}_i$, $\boldsymbol{\theta}_i$ (i.e. the MZI phase shifts) until each $k$\textsuperscript{th} output produced matches the $k$\textsuperscript{th} target state $\left|\mathrm{targ}\right\rangle_k$. Mathematically, this involves minimizing the cost function,
\begin{equation} \label{eq:cost}
   \mathcal{C} = 1 - \frac{1}{K}\sum_{k=1}^K\left|{}_k\!\left\langle\mathrm{targ}\right|\mathbf{S}\left|\mathrm{in}\right\rangle\!{}_k\right|^2,
\end{equation}
a measure of the network error evaluated by comparing all $K$ output states to the corresponding targets. The fidelity of a QPNN trained to perform a mapping between $K$ input ($\left|\mathrm{in}\right\rangle$) and target ($\left|\mathrm{targ}\right\rangle$) state pairs is given by,
\begin{equation} \label{eq:fidelity}
    \mathcal{F} = \frac{1}{K}\sum_{k=1}^K \frac{\left|{}_k\!\left\langle\mathrm{targ}\right|\mathbf{S}\left|\mathrm{in}\right\rangle\!{}_k\right|^2}{\sum_{\left|x\right\rangle\in \mathrm{CB}} \left|\left\langle x\right|\mathbf{S}\left|\mathrm{in}\right\rangle\!{}_k\right|^2},
\end{equation}
where $\mathrm{CB}$ is the set of computational basis states, all those that abide by the logical dual-rail encoding of the photonic qubits. Simply put, the fidelity quantifies the similarity between the actual output and the target state, if no photons have been lost and the output remains dual-rail encoded.\\

\noindent\textbf{Performing Network Simulations}\\
All QPNN simulations were facilitated by the \texttt{quotonic} (v1.0.0) package, a \texttt{python} (v3.10.2) framework that we designed to efficiently model and train the networks. This package is inspired by previous work on QPNNs in Refs.~\cite{Steinbrecher:19, Ewaniuk:23, Basani:24}. It relies on \texttt{jax} (v0.4.30) for numerical computation and \texttt{optax} (v0.2.3), with the default Adam optimizer and exponential decay scheduler, to perform each optimization routine. Given that Adam is a gradient-based optimization algorithm \cite{Kingma:17}, we use the version of \texttt{autograd} native to \texttt{jax} to compute analytical gradients of the cost function (Eq.~\ref{eq:cost}) while training the network. Each optimization trial begins by selecting random linear unitary transformations from the Haar measure, for each layer, and performing Clements decomposition \cite{Clements:16} to extract the initial phase shift parameters $\boldsymbol{\phi}_i$, $\boldsymbol{\theta}_i$, as this has been shown to improve convergence speed \cite{Pai:19}. Training proceeds for a set number of epochs that was calibrated empirically for each model. All simulations were conducted on the Frontenac Platform computing cluster offered by the Centre for Advanced Computing at Queen's University.\\

\noindent\textbf{Modeling Network Imperfections}\\
To model imperfections in the QPNN, we follow the procedure of Ref.~\cite{Ewaniuk:23} such that each individual component introduces its own amount of loss or routing errors, selected from a distribution, as is typically the case experimentally \cite{Harris:18}. Routing errors arise from imbalance in the nominally $50:50$ directional couplers (DCs) that form each MZI, which we calculate as $0.5\%$ in Supplementary Sec.~S7 from the results quoted in Ref.~\cite{Alexander:25}, the \textit{single} platform. It follows that the splitting ratio for each DC is selected from a normal distribution centered at 0.5 with a width of 0.005. Similarly, the loss introduced by each MZI is sampled from a normal distribution with a mean (standard deviation) of 0.2130 (0.0124) dB, 0.0210 (0.0016) dB, 0.00210 (0.00016) dB for the \textit{single}, \textit{multi}, and \textit{future} models, respectively.\\

\noindent\textbf{Calculating Tree State Generator Metrics}\\
Trees are defined by branching vectors $\vec{b}$ (cf. Fig.~\ref{fig:generator}b) of length $d$, where $d$ is the tree depth. Starting from the root photon, residing in row 0 at the top, each element $b_j$ specifies the number of branches stemming from each photon in row $j$ to row $j + 1$. To meet the demands of the generation protocol, the time delay enacted by the $i$\textsuperscript{th} delay line while emitting photons in row $j$ of the tree is given by,
\begin{equation} \label{eq:delay}
   \Delta t_i^{(j)} = \left(\prod_{k=0}^{d-1} b_k - (i - 1)\prod_{k=j}^{d-1}b_k\right)\Delta t_s,
\end{equation}
where $b_d\equiv 1$, and delay lines are numbered from $i = 1$ to $i = b_{j - 1}$. The first delay line, $\Delta t_1^{(j)}$, is the same for all $j$ because it is reused in each of the $d$ stages of the protocol. In fact, the total time required to generate a tree, 
\begin{equation} \label{eq:total-time}
   \Delta t_T = \left(\prod_{k=0}^{d-1}b_k\right)d\Delta t_s,
\end{equation}
can also be expressed as $d\Delta t_1^{(j)}$. To convert from time to distance, we assume the use of SMF-28 optical fibers that have 0.17 dB/km loss and a group index of 1.462 at 1550 nm \cite{Corning}. If the physical delay lines are static, then the maximum number of lines required is $\sum_{k=0}^{d-1}b_k$, assuming that none can be reused in different steps of the protocol. If they are dynamic, this number is reduced to $\mathrm{max}\{\vec{b}\}$. For $N_d$ physical delay lines, regardless of whether they are dynamic, the switch at the output of the QPNN must be $1\times(N_d+1)$. If the delay lines are static, then a $1\times(N_d-1)$ switch is also required at the input of the QPNN. Following the formalism of Ref.~\cite{Varnava:06}, a tree with branching vector $\vec{b}$ contains,
\begin{equation} \label{eq:photons}
    n = 1 + \sum_{j=0}^{d-1}\prod_{k=0}^{j}b_k,
\end{equation}
total photons that together encode one logical qubit. When each individual photon has a probability $\epsilon_0$ of being lost, the effective loss (i.e. the probability that the information is lost) is given by,
\begin{equation} \label{eq:effective-loss}
    \epsilon_\mathrm{eff} = 1 - (1 - \epsilon_0)P_\mathrm{ind},
\end{equation}
which combines the probability that the root photon survives ($1-\epsilon_0$) with the probability that its state can be recovered via indirect measurement, $P_\mathrm{ind}$. The latter is calculated using a recursive relation,
\begin{equation} \label{eq:p-indirect}
    P_\mathrm{ind} = \left[\left(1-\epsilon_0+\epsilon_0R_1\right)^{b_0} - (\epsilon_0R_1)^{b_0}\right]\left(1-\epsilon_0+\epsilon_0R_2\right)^{b_1},
\end{equation}
where $R_j$ is the probability of a successful indirect measurement on a photon in the $j$\textsuperscript{th} row of the tree, expressed as,
\begin{equation}
    R_j = 1 - \left[1-\left(1-\epsilon_0\right)\left(1-\epsilon_0+\epsilon_0R_{j+2}\right)^{b_{j+1}}\right]^{b_j},
\end{equation}
for $j < d$, with $R_d\equiv0$, $b_d\equiv0$ taken as the initial values. In both Figs.~\ref{fig:generator}e-f and Fig.~\ref{fig:boosting}, we performed a search over all branching vectors with $\max\{\vec{b}\}\leq 4$ up to maximum depths of 8 and 6, respectively, and evaluated $\epsilon_\mathrm{eff}$ for each. The branching vectors were constrained as such since we could only simulate QPNNs up to $10\times 10$, operating on 5 dual-rail photonic qubits, in a reasonable amount of time. That being said, it is clear from the results that this search domain well-encapsulates the optimal tree shapes for the analysis of Fig.~\ref{fig:boosting}. With $\epsilon_\mathrm{eff}$ defined, the communication rate in Fig.~\ref{fig:boosting}d is simply the product of the effective transmission ($1-\epsilon_\mathrm{eff}$) with the repetition rate ($1/\Delta t_T$). In contrast, the generation rate (cf. Fig.~\ref{fig:growing}b) is the product of the joint probability that each photon survives the generator with the repetition rate. If each of the $n$ photons in the tree experiences the same amount of loss in the generator, $\epsilon_0$, then this joint probability is simply $(1-\epsilon_0)^n$. However, we computationally model the different amounts of loss experienced by each individual photon in each tree to more accurately describe the generation rate.


\section{Data Availability} \label{sec:data}
The code written to produce the findings of this study, namely the \texttt{quotonic} package, is available at \url{https://github.com/jewaniuk/quotonic/}. The data produced during this work are openly available in the Borealis repository of the Queen's University Dataverse at \url{https://doi.org/10.5683/SP3/RNNOGK}.


\section{Acknowledgments} \label{sec:acknowledgments}
This research was supported by the Vector Scholarship in Artificial Intelligence, provided through the Vector Institute. The authors thank T. Schröder for his insights on tree-type photonic cluster state generation and gratefully acknowledge the support of the Natural Sciences and Engineering Research Council of Canada (NSERC), the Canadian Foundation for Innovation (CFI), and Queen's University.


\section{Author Contributions} \label{sec:contributions}
N.R. and J.E. conceived the idea for a tree state generator based on QPNNs. J.E. developed the generation protocol and used it to perform all simulations and analysis, with supervision from B.S. and N.R. The results were discussed by all authors, who also shared the writing and editing responsibilities for the manuscript.

\bibliography{references}

\end{document}


\title{Supplementary Information for ``Large-Scale Tree-Type Photonic Cluster State Generation with Recurrent Quantum Photonic Neural Networks''}

\author{Jacob Ewaniuk}
\email{jacob.ewaniuk@queensu.ca}
\affiliation{Centre for Nanophotonics, Department of Physics, Engineering Physics \& Astronomy, 64 Bader Lane, Queen's University, Kingston, Ontario, Canada K7L 3N6}

\author{Bhavin J. Shastri}
\email{bhavin.shastri@queensu.ca}
\affiliation{Centre for Nanophotonics, Department of Physics, Engineering Physics \& Astronomy, 64 Bader Lane, Queen's University, Kingston, Ontario, Canada K7L 3N6}
\affiliation{Vector Institute, Toronto, Ontario, Canada, M5G 1M1}

\author{Nir Rotenberg}
\email{nir.rotenberg@queensu.ca}
\affiliation{Centre for Nanophotonics, Department of Physics, Engineering Physics \& Astronomy, 64 Bader Lane, Queen's University, Kingston, Ontario, Canada K7L 3N6}

\date{\today}

\maketitle

\section{Comparison of Tree-Type Photonic Cluster State Generation Protocols} \label{sec:comparison}
In Figs.~1e-f, we compare our QPNN-based protocol with the linear-optical (lo, blue) protocol of Ref.~\cite{Bodiya:06} and the emitter-based protocol of Ref.~\cite{Zhan:20}, when quantum dots (qd, purple), silicon-vacancy centers in diamond (SiV, orange), and atoms (red) are used. In this section, we will explain how each effective loss (cf. Fig.~1e) and repetition rate (cf. Fig~1f) was calculated for the latter protocols, given that the main text explains how these measures are computed for our QPNN-based one.

Considering the linear-optical protocol first, it is evident that its effective loss overlaps identically with the QPNN-based protocol in Fig.~1e. This is a result of the fact that the equations are the same. Since the linear-optical protocol is not hindered, at any scale, by decoherence during generation, the effective loss is given simply by Eq.~7. However, the repetition rate is drastically different since the linear-optical protocol relies on probabilistic operations that lead to exponentially increasing failure rates as the trees scale. Fortunately, the authors of Ref.~\cite{Bodiya:06} give both exact and approximate expressions for the total preparation time $\Delta t_T$ in their Eq.~5, which can be computed from the recursive relation of their Eq.~4 alongside the success probability for each probabilistic operation, given in their Eq.~3. To make a fair comparison, we assume the linear-optical protocol has perfect source and detector efficiencies, and set the source rate of the protocol as $1/\Delta t_s= 100$~MHz, the same as that used for the QPNN-based protocol. With the total preparation time, the repetition rate is simply calculated as $1/\Delta t_T$. We use the approximate expression in Eq.~5 of Ref.~\cite{Bodiya:06}, valid for $n>>1/2$ when the detectors are perfect, when the number of photons $n > 100$ since it is easier to calculate and allows us to easily match the scales of the trees for a more accurate comparison.

For the emitter-based protocol, we use Eqs.~S1-S4 from Ref.~\cite{Zhan:20} for Fig.~1e, which together provide a correction to the effective loss of a tree when decoherence effects are included. To calculate these equations, we must specify the source rate ($1/\Delta t_s$) as well as the coherence time of the emitter ($t_\mathrm{coh}$). Since this protocol uses the emitter as its source of photons, it is necessary to consider the typical tradeoff between emitter coherence times and their lifetimes ($1/\gamma_L$). In other words, emitters that remain coherent for longer also tend to take longer to emit photons, thus reducing the source rate. It is for this reason that we add different curves to Figs.~1e-f for different kinds of emitters. We specifically follow the recommendations made by the authors of Ref.~\cite{Zhan:20}, selecting the same combinations of parameters that they quote as the state-of-the-art for each emitter type. These parameters are summarized in Tab.~\ref{tab:emitter} below, with references to the original experiments that produced each result.
\begin{table}
\caption[Experimental parameters used to evaluate the emitter-based protocol of Ref.~\cite{Zhan:20}]{Each parameter is quoted as the state-of-the-art by the authors of Ref.~\cite{Zhan:20}, and the corresponding references to where these results were shown are included. This is done for each quantum emitter considered in Figs.~1e-f of the main text: quantum dots (qd), silicon vacancy centers in diamond (SiV), and atoms.}
\begin{ruledtabular}
\begin{tabular}{ccccc}
     Emitter & CZ Gate Bandwidth, $\gamma_R$ [GHz] & Ref. & Emitter Coherence Time, $t_\mathrm{coh}$ [ms] & Ref.\\
    \hline
    qd & $2\pi\times80$ & \cite{Ota:18} & 0.004 & \cite{Huthmacher:18}\\
    SiV & $2\pi\times0.1$ & \cite{Bhaskar:20} & 10 & \cite{Sukachev:17}\\
    atom & $2\pi\times10$ & \cite{Brekenfeld:20} & 1000 & \cite{Reiserer:15}\\
\end{tabular}
\end{ruledtabular}
\label{tab:emitter}
\end{table}
Here, $\gamma_R$ is the bandwidth of the emitter-photon scattering CZ gate applied in this emitter-based protocol. The authors require $\gamma_L=0.001\gamma_R$ to ensure high-fidelity operation, and state that the time dedicated to each individual photon (i.e. equivalent to what we denote as the source time $\Delta t_s$ here) is $\Delta t_s = 6/\gamma_L$. These two relations allow us to translate the experimental value for $\gamma_R$ to $\Delta t_s$ as required to evaluate both the effective loss and repetition rate. On that note, though the authors do not give a general expression for the total preparation time, it is trivial to derive by following the protocol they describe. For reference, we derived it as,
\begin{equation} \label{eq:total-time-emitter}
    \Delta t_T = \left(\prod_{k=0}^{d-1}b_k + \prod_{k=0}^{d-2}b_k - 1\right)\Delta t_s + d\left(\prod_{k=0}^{d-1}b_k + \prod_{k=0}^{d-2}b_k\right)\Delta t_s,
\end{equation}
where $d$ is the tree depth, the length of branching vector $\vec{b}$ with elements $b_k$, as in the main text. Again, the repetition rate follows as $1/\Delta t_T$.

To produce the curves shown in Figs.~1e-f, we iterate from a tree depth of 1 to 8. For each depth, we perform a search of all possible branching vectors where $\max\{\vec{b}\} \leq 4$, and add markers to each curve if the calculated effective loss (in the absence of decoherence) is reduced from the previous minimum. This allows us to isolate the general performance of each protocol as the trees scale without clouding the results with specific tree shapes that tend to perform worse than similarly-scaled ones. As stated in the main text, we assume for all protocols that each individual photon experiences $10\%$ loss during generation followed by $\sim18\%$ loss due to the 5 km fiber channel. Although the loss during generation would vary between the protocols, all suffer loss in similar forms, so giving them each the same amount still allows for a fair comparison.

\section{Additional Analysis from Training QPNNs} \label{sec:training}
\begin{figure}[ht]
\centering
\includegraphics{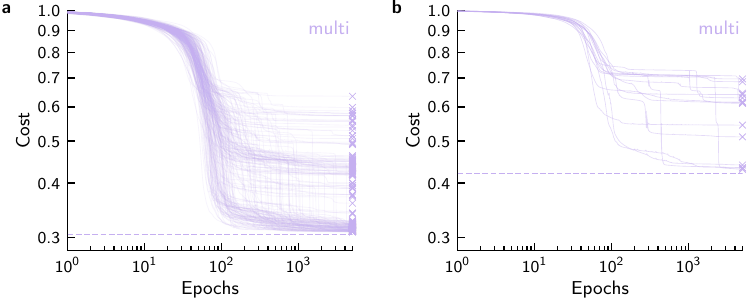}
\caption[Training QPNNs to generate tree unit cells with increased branching]{\textbf{a} (\textbf{b}) Minimization of the network cost (i.e. average error) during 200 (15) optimization trials of 5000 epochs each for the \textit{multi}-platform (see main text for platform details). Dashed lines denote the loss limit (i.e. minimum achievable cost due to loss). In these trials, the directional coupler splitting ratios are $(50\pm5)\%$ such that there are increased routing errors as compared to the results shown in the main text (where they are 0.5\%).}
\label{fig:training-larger}
\end{figure}

In the main text, we demonstrate that the QPNN can learn to perform all operations required to support the generation of tree unit cells with two branches (i.e. $b = 2$). However, the QPNN-based approach is not limited to this branching ratio, as will be described further in Sec.~\ref{sec:general-protocol}. Here, we show results for larger QPNNs trained to accommodate higher branching ratios. Specifically, in Fig.~\ref{fig:training-larger}a (b) we train 200 (15) QPNNs in 5000 epochs each to accommodate all unit cells that have branching $b_k\leq 3$ (4).
In these optimization trials, we assume the \textit{multi}-platform alone for proof-of-principle, and since the simulations become much more computationally intensive as the QPNNs scale. In particular, we could only run 15 trials for Fig.~\ref{fig:training-larger}b since the QPNNs must be $10\times 10$, operating on up to 5 photons simultaneously, to accommodate unit cells with a branching ratio of 4. With that said, we find that in both cases the QPNN achieves loss-limited operation, and would expect this trend to continue as the branching, and QPNN, scale further. It is worth noting, however, that the number of layers in the QPNN was forced to increase from 2 for $b=2$, as shown in the main text, to 3 layers here, for both $b_k\leq 3$ and 4.

\begin{figure}[ht]
\centering
\includegraphics{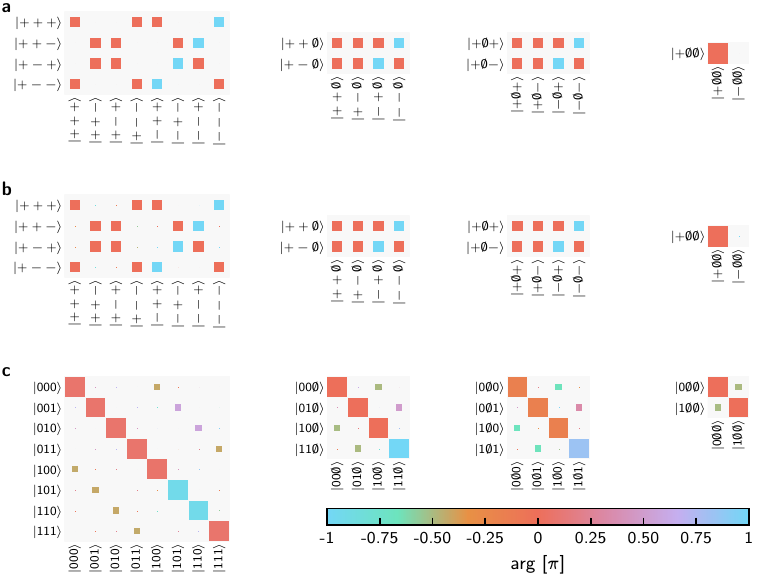}
\caption[Comparison of Hinton diagrams for the QPNN operations required for 2-branch-only trees]{\textbf{a} (\textbf{b}) Ideal (resultant) Hinton diagrams resolved in the X-basis for each operation of the \textit{multi}-platform QPNN outlined in black in Fig.~2e, where the uppermost photonic qubit is $\left|+\right\rangle$ at the input (vertical axis), yet can belong to a superposition of $\left|+\right\rangle$ and $\left|-\right\rangle$ at the output (horizontal axis). \textbf{c} Resultant Hinton diagram resolved in the Z-basis for each operation of the \textit{multi}-platform QPNN outlined in black in Fig.~2e. Each box is colored according to its argument, as shown on the colorbar. When a photonic qubit is missing at any input or output port of the network, $\emptyset$ is written in its place.}
\label{fig:hinton-comparison}
\end{figure}

Next, we show additional Hinton diagrams for all QPNN operations required to generate trees with $b = 2$ (cf. Figs.~2a-d), as discussed in the main text. First, in Fig.~\ref{fig:hinton-comparison}a,
we show the ideal Hinton diagrams for each operation in the X-basis, again where the vertical (horizontal) axis shows input (output) states. When we compare these diagrams with those included in the main text for the best \textit{multi}-model trial, which are copied here to Fig.~\ref{fig:hinton-comparison}b for reference, we see that there is effectively no difference between them. This is what we would expect given the near-unity fidelities that the QPNN achieves for each individual operation (cf. Figs.~2f-i). In Fig.~\ref{fig:hinton-comparison}c, we show the Hinton diagrams for the same \textit{multi}-model trial, yet this time in the Z-basis. Immediately, it appears that the diagrams do not exactly match what would be expected for the multi-CZ-gate circuits that we assume the QPNN should form to fulfill its role in the tree generation protocol. This is a result of the chosen training set used during network optimization. To meet all criteria for the tree generator, the QPNN must entangle the newly emitted photon in each timestep with any and all that arrive simultaneously by applying CZ gates between them, regardless of how many photons there are. If the newly emitted photon enters alone, the QPNN performs an identity operation instead. However, it is clear from the architecture of the generator (cf. Fig.~1a) that the newly emitted photon will always enter the QPNN in state $\left|+\right\rangle$, regardless of the timestep. Therefore, we improved simulation efficiency by reducing the size of the training set, training on only the subset of input-target state pairs where the first photon is in state $\left|+\right\rangle$. Not only does this choice reduce the number of calculations per epoch, it also widens the solution space during optimization. This can be seen from Fig.~\ref{fig:hinton-comparison}c, where the nonzero amplitudes and abnormal phases of the off-diagonal elements are not erroneous. Instead, they cancel each other out when considering the subset of operations required by the tree generator. This is easiest to understand by considering the identity operation. If an equal superposition of the two input states (on the vertical axis, i.e. a $\left|+\right\rangle$ state) is sent to the QPNN, the resultant output will be unchanged.

\begin{figure*}[ht!]
\centering
\includegraphics{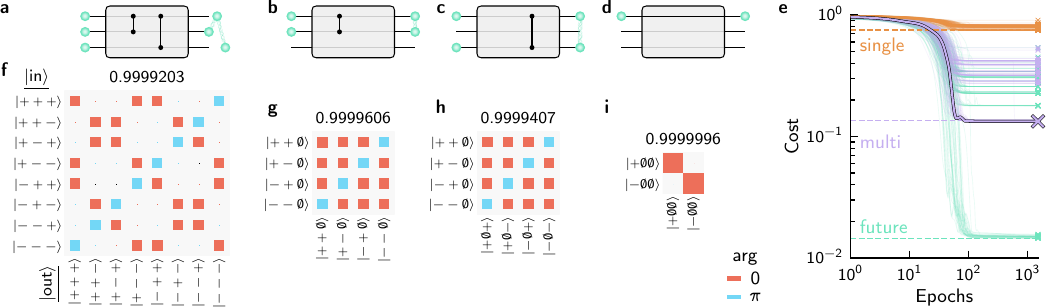}
\caption[Training a 2-layer, 6-mode QPNN to realize a general two-CZ-gate logical circuit]{\textbf{a}-\textbf{d} Circuit diagrams for each QPNN operation required by the generation protocol. \textbf{e} Minimization of the network cost (i.e. average error) during 200 optimization trials of 1500 epochs each for the \textit{single}-, \textit{multi}- and \textit{future}-platform (see main text for platform details). Dashed lines denote the loss limit (i.e. minimum achievable cost due to loss). \textbf{f}-\textbf{i} Hinton diagrams resolved in the X-basis for each operation of the \textit{multi}-platform QPNN outlined in black in \textbf{e}, where the input (output) states are shown on the vertical (horizontal) axis. Each box is colored according to its argument, which is always within $\pi/100$ of either $0$ or $\pi$ up to an insignificant global phase. When a photonic qubit is missing at any input or output port of the network, $\emptyset$ is written in its place. The fidelity of each operation is given above its Hinton diagram, never falling below 0.9999203 for this network.}
\label{fig:training-full}
\end{figure*}

Based on the previous discussion, we now clarify exactly what forms the training set for the QPNN in our simulations. Specifically, we choose all input states in the X- and Z-bases where the first photon is in state $\left|+\right\rangle$, then compute the corresponding target states by applying the multi-CZ circuit as required for cluster state generation. If it were necessary to fully generalize the QPNN circuit to match the multi-CZ-gate logical circuit that would typically be envisioned for cluster state generation, this can still be achieved at the cost of slightly increased training complexity. In Fig.~\ref{fig:training-full}, we show a duplicate of Fig.~2 from the main text where the QPNN is instead trained on every possible input in both the X- and Z-bases.
Each operation still achieves near-unity fidelity for the selected \textit{multi}-model QPNN, and QPNNs from each model still achieve loss-limited performance. This kind of circuit generalization could become necessary when extending the QPNN-based protocol to other cluster state types or adapting the generator architecture to boost generation rates.

\section{Generalized QPNN-Based Generation Protocol} \label{sec:general-protocol}
\begin{figure}[t]
\centering
\includegraphics[scale=0.82]{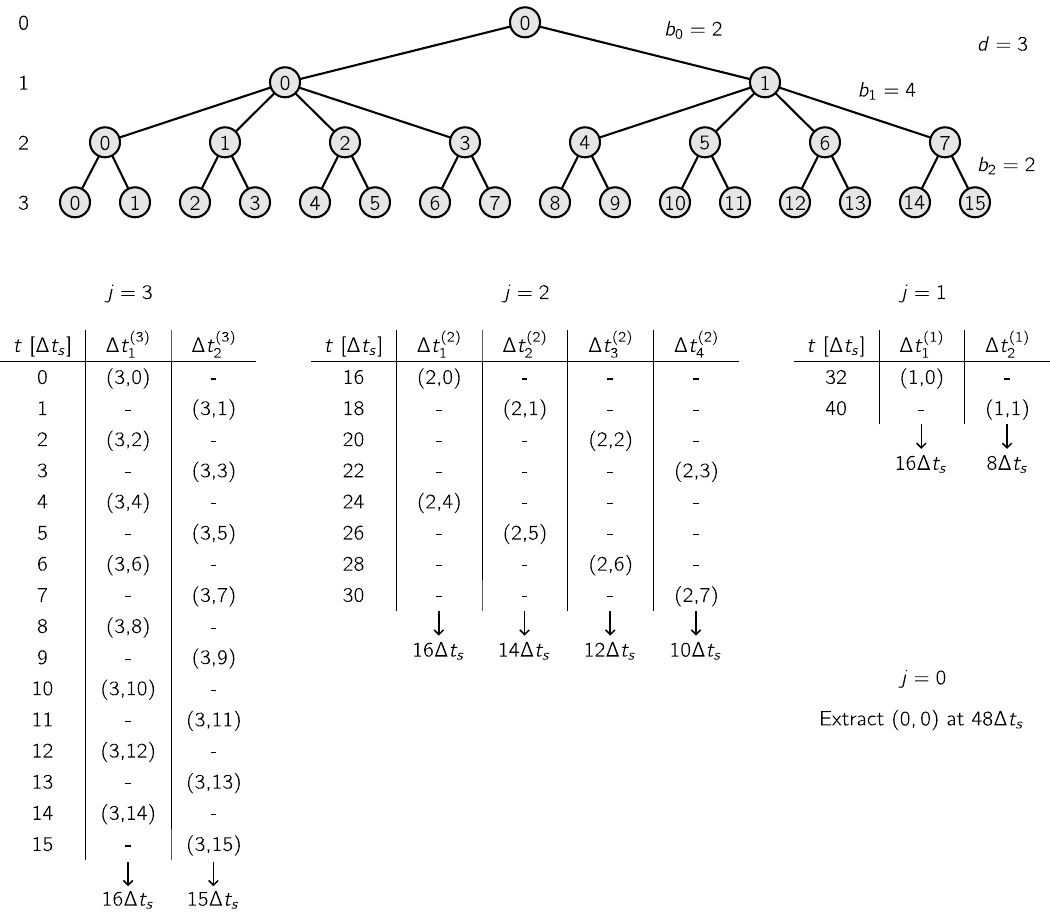}
\caption[Example of the generalized QPNN-based tree state generation protocol]{Protocol for generating a tree-type photonic cluster state with branching vector $\vec{b}=[2,4,2]$ (depth $d=3$) using our QPNN-based approach. The tree is drawn and labeled at the top for reference. Below it, the tables denote which photon is emitted at each timestep and which delay line they should be routed to after traversing through the QPNN. They are separated by $j$, which is the row of the tree being generated, starting from the bottom and working upward. The entire protocol lasts 48$\Delta t_s$, where $\Delta t_s$ is the time dedicated to the emission of an individual photon by the source.}
\label{fig:protocol-b242}
\end{figure}
Here, we provide a general timing protocol for generating tree-type photonic cluster states of arbitrary branching vectors, $\vec{b}$, using the architecture described in the main text (cf. Fig.~1a-d). The protocol always constructs the tree from the bottom (row $j = d$, where $d$ is the tree depth) to the top (row $j = 0$). Each stage of the protocol corresponds to the $j$\textsuperscript{th} row of the tree, including all timesteps where a photon in that row is emitted by the single-photon source. In describing this protocol, we assume $b_d \equiv 1$ even though $b_d$ is not actually defined in the branching vector $\vec{b}$.
\begin{enumerate}
    \item There are $\prod_{k=0}^{d-1}b_k$ photons in the bottom row ($j = d$) of the tree. Emit each of these photons in subsequent timesteps $t$, separated by $\Delta t_s$, the time allotted to the source. Each emitted photon enters the QPNN alone such that the network applies an identity operation where the photon is simply routed through to its output. Following the QPNN, the photon is routed by the switch to one of $b_{d-1}$ delay lines, starting at line $i = 1$ ($\Delta t_{1}^{(d)}$), incrementing each timestep to line $i = b_{d-1}$ ($\Delta t_{b_{d-1}}^{(d)}$), then repeating. As calculated from Eq. 1 of the main text, each subsequent delay line is $\Delta t_s$ shorter than the previous.
    \item For each $j$\textsuperscript{th} row of tree, from $j = d-1$ to $j = 1$, emit a new photon at the single-photon source in intervals of $\left(\prod_{k=j}^{d}b_k\right)\Delta t_s$. This parent photon will then arrive with its $b_j$ children photons at the input of the QPNN such that the network entangles them according to the target tree shape (i.e. a CZ gate operation is performed between the parent and each child). All children photons are routed to the output of the generator. The parent photon is routed by the switch to one of $b_{j-1}$ delay lines, starting at $i = 1$ ($\Delta t_{1}^{(j)}$), incrementing each $\left(\prod_{k=j}^{d}b_k\right)\Delta t_s$ interval to line $i = b_{j-1}$ ($\Delta t_{b_{j-1}}^{(j)}$), then repeating. As calculated from Eq. 1 of the main text, each subsequent delay line is $\left(\prod_{k=j}^{d}b_k\right)\Delta t_s$ shorter than the previous.
    \item At the top row ($j = 0$) of the tree, the root photon is emitted to join its $b_0$ children at the input of the QPNN, where again, the QPNN entangles them according to the target tree shape. Rather than routing the root photon through an unnecessary delay line, the switch routes it to the output of the generator with its children.
\end{enumerate}

Over the course of the protocol, the maximum number of photons that the QPNN must be able to simultaneously operate on is $\max\{\vec{b}\} + 1$. Therefore, the QPNN must consist of at least $L$ layers of $2(\max\{\vec{b}\} + 1) \times 2(\max\{\vec{b}\} + 1)$ meshes of Mach-Zehnder interferometers in the dual-rail encoding scheme. At a minimum, $L = 2$, which we found to be sufficient for the case of $\max\{\vec{b}\}=2$ as demonstrated in the main text. However, we empirically discovered that $L = 3$ was required for both $\max\{\vec{b}\}=3$ and $\max\{\vec{b}\}=4$ (cf. Fig.~\ref{fig:training-larger}). This suggests that the scaling of the network layers is sublinear with $\max\{\vec{b}\}$, yet we could only verify this for up to $\max\{\vec{b}\}=4$ given our current simulation capabilities.

While all of the steps outlined above apply regardless of whether the delay lines are dynamic, there is an additional switch required if they are static. This switch connects all delay lines, except the longest one, to a subset of the inputs of the QPNN, such that delay lines of different lengths can be chosen at timesteps dictated by the protocol above. Specifically, for $N_d$ physical static delay lines, this switch should be $(N_d-1)\times(\max\{\vec{b}\}-1)$, given that it maps all delay lines except the longest to all inputs of the QPNN except those that connect to the source and the longest delay line, respectively. The input-output mapping of this switch should be updated before generating each $j$\textsuperscript{th} row of the tree to select the delay lines $\Delta t_i^{(j)}$ required.

As an example of the general protocol, Fig.~\ref{fig:protocol-b242} denotes each timestep $t$ when a photon should be emitted at the single-photon source to form a tree with branching vector $\vec{b} = [2, 4, 2]$. The relevant delay line lengths, as calculated from Eq. 1 of the main text, are annotated for reference. As always, the protocol begins by sequentially emitting each photon along the bottom row of the tree, and routing them to enough delay lines to match the final element of the branching vector, $b_{d-1}$. After each row of the tree is generated, the required delay lines are adjusted to conform to the spacing between the remaining photons. In the second row, four delay lines are used to ensure that all four children photon will reach the QPNN together when their parent is subsequently emitted. Altogether, the entire protocol lasts $48\Delta t_s$.

\section{Modeling the Tree State Generator \& Evaluating its Performance} \label{sec:generator}
To reiterate the main text, the \textit{single}-platform model entirely considers the results quoted in Ref.~\cite{Alexander:25}, even once the losses at the switches and chip-to-fiber couplers are included when describing the operation of the generator as a whole. The \textit{multi}-platform model ties in state-of-the-art results demonstrated in separate experiments and platforms, to give an indication of what the integration of current photonic elements would provide. Then, the \textit{future}-platform improves the loss of each photonic element by one order-of-magnitude to provide a look ahead. In all models, the fiber losses are held constant, considering ultra-low-loss telecom SMF-28 fibers. All loss values are summarized in Tab.~\ref{tab:loss} with corresponding references to where these losses are reported.
\begin{table}[h]
\caption[Losses for each experimental model considered in the main text]{For each model, we specify the loss value for each photonic element (DC: directional coupler, PS: phase shifter) in the QPNN-based tree state generator, as well as the corresponding references where these values are taken from (other than the \textit{future} model which is a projection).}
\begin{ruledtabular}
\begin{tabular}{ccccccccc}
     Model & DC Loss [dB] & Ref. & PS Loss [dB] & Ref. & Switch Loss [dB/stage] & Ref. & Coupling Loss [dB] & Ref. \\
    \hline
    \textit{single} & 0.0005 & \cite{Alexander:25} & 0.106 & \cite{Alexander:25} & 0.107 & \cite{Alexander:25} & 0.120 & \cite{Alexander:25}\\
    \textit{multi} & 0.0005 & \cite{Alexander:25} & 0.010 & \cite{Parra:20, Harris:14} & 0.061 & \cite{Alexander:25, Eltes:19, Krasnokutska:18, Nozawa:91} & 0.120 & \cite{Alexander:25}\\
    \textit{future} & 0.00005 & - & 0.001 & - & 0.0061 & - & 0.0120 & -\\
\end{tabular}
\end{ruledtabular}
\label{tab:loss}
\end{table}
\begin{figure}[b]
\centering
\includegraphics{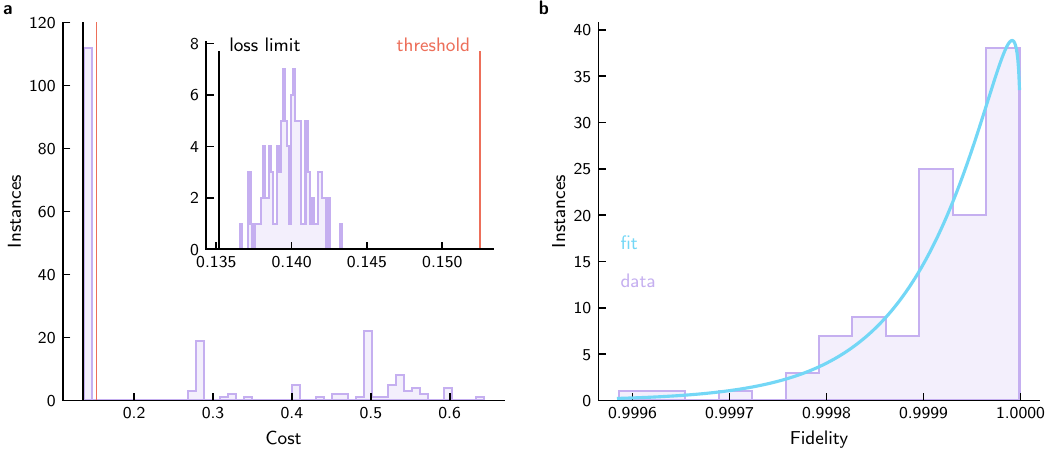}
\caption[Statistical analysis of trained QPNN fidelity]{\textbf{a} Histogram of the optimized cost (cf. Eq.~2) for \textit{multi}-model QPNNs trained on all operations required for generating trees with $b = 2$, as considered in the main text (see Fig.~2 for full training data). Successful optimizations achieve near-loss-limited performance, where the loss limit is drawn as a black line. To select these trials for further analysis, we define a threshold (red line) to eliminate the failed trials. This is more clearly seen in the inset, which zooms in on the successful trials. \textbf{b} Histogram of the network fidelity for all successful trials isolated in \textbf{a}, alongside a beta distribution fit which is used to extract a mean fidelity and corresponding 95\% confidence intervals.}
\label{fig:statistics}
\end{figure}
From these values, the loss per MZI (switch stage) is calculated by summing that for two directional couplers and two (one) phase shifters. Since the switches are simply routing the photons from one port to another, they can be modeled using cascaded stages of single-phase-shifter MZIs. For $N_d$ physical delay lines, the switch at the output of the QPNN must have $1\times(N_d+1)$ ports ($2\times(2N_d+2)$ spatial modes), thus requiring $\lceil\log_2\left(2N_d+2\right)\rceil - 1$ stages of single-phase-shifter MZIs. If the delay lines are static, then a switch at the input of the QPNN is also necessary, with $1\times(N_d-1)$ ports ($2\times(2N_d-2)$ spatial modes) and $\lceil\log_2\left(2N_d-2\right)\rceil - 1$ stages of single-phase-shifter MZIs. These relations allow us to calculate the switch loss regardless of the tree shape. Finally, it is worth noting how we landed on the value of 0.061 dB/stage switch loss for the \textit{multi} model. Recall that the switch is required to operate at 100 MHz, so it cannot rely on thermo-optic phase shifters like the MZIs of the QPNN, but can be realized using electro-optic phase shifters like those fabricated with barium titanate or lithium niobate. The electro-optic phase shifter demonstrated in Ref.~\cite{Alexander:25} has 53 dB/m loss and is 2 mm long. Thus, one would estimate the loss of a 1 mm long phase shifter in the same platform as 0.053 dB, at the expense of increasing the operating voltage by a factor of two. Electro-optic phase shifters using the same material (barium titanate) were fabricated with 1 mm length in Ref.~\cite{Eltes:19}. Additionally, lithium niobate waveguides have been fabricated with losses of 40 dB/m \cite{Krasnokutska:18} and 20 dB/m \cite{Nozawa:91}, suggesting losses of 0.08 dB and 0.04 dB, respectively, if the phase shifters were made to be 2 mm in length. With all of this considered together, we take 0.06 dB loss per electro-optic phase shifter as a fair estimate for the \textit{multi} model, which leads to 0.061 dB per switch stage when combined with two directional couplers.

Next, we describe the statistical analysis performed to extract the QPNN-based generator fidelities shown in Fig.~4a. Taking the \textit{multi}-model training results from Fig.~2e, as an example, we plot a histogram of the optimized cost values achieved over the 200 trials in Fig.~\ref{fig:statistics}a. Here, it is evident that the successful optimization trials all fall within a single bin near the loss limit (black line), which we zoom in on in the inset. We must first isolate these successful trials before characterizing the network fidelity, given that in a practical scenario, training would be repeated until the optimization is successful (i.e. near-loss-limited). If the loss limit is expressed as $\mathcal{C}_{\ell\ell}$, then we define the threshold as $1-0.90(1-\mathcal{C}_{\ell\ell})$, $1-0.98(1-\mathcal{C}_{\ell\ell})$, and $1-0.98(1-\mathcal{C}_{\ell\ell})$ for the \textit{single}-, \textit{multi}-, and \textit{future}-model QPNNs, respectively. These thresholds were empirically set to isolate the successful trials for each model, as shown in the inset of Fig.~\ref{fig:statistics}a. After thresholding the optimization trials, we compute the optimized fidelity for each successful one and form a histogram of the results as in Fig.~\ref{fig:statistics}b. This histogram is fit with a beta distribution, which is particularly suitable for describing probability distributions of probabilities (fidelity is the probability that the correct output is generated when it is logical). From this fit, we can extract the mean fidelity and corresponding 95\% confidence intervals to define $\mathcal{F}$ with errors. Since the QPNN is applied $n$ times when generating a tree of $n$ photons, the overall tree fidelity can be calculated as $\mathcal{F}^n$, and we propagate the errors to match.

\begin{figure}[ht]
\centering
\includegraphics{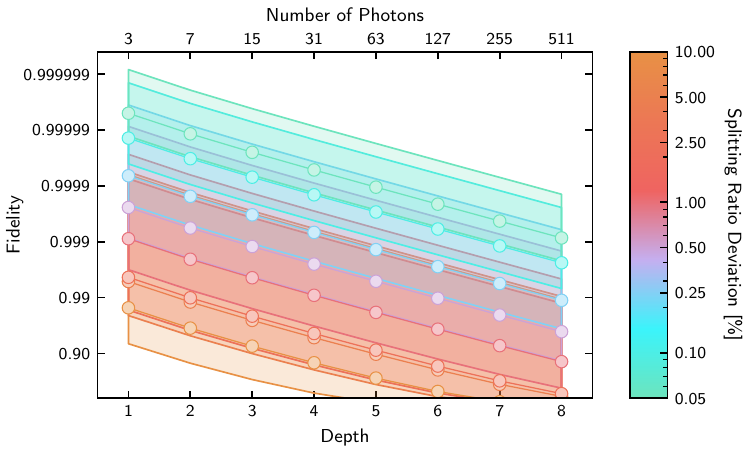}
\caption[Tree fidelity as the directional coupler splitting ratio deviation is swept]{This analysis mimics the fidelity shown in Fig.~4 from the main text, using the \textit{multi}-model QPNN with different levels of directional coupler splitting ratio imbalance. For example, if the deviation is 5\%, then each splitting ratio for each directional coupler is sampled from a normal distribution centered at 50\% with a width of 5\%. In the main text, we consider the state-of-the-art ratio of 0.5\%. The markers (shaded regions) denote the mean (95\% confidence intervals) of fit beta distributions. Each point is generated from data gathered by training 200 QPNNs, each in 1000 epochs, as in the main text.}
\label{fig:dc-sweep}
\end{figure}

In the main text, we found that the fidelity of the tree generator was limited by the imperfect routing of the directional couplers that form the MZIs, rather than unbalanced loss. Here, we extend this analysis by varying the directional coupler splitting ratio variation, and plot the fidelity as a function of tree depth in Fig.~\ref{fig:dc-sweep}.
From this sweep, it is clear that imperfect routing is the dominant limitation to the fidelity of the tree generator. The fidelity steadily increases as the splitting ratios become more balanced. By halving the deviation from the current best, to 0.25\% (blue) from 0.5\% (purple), the generation of 511-photon trees will approximately reach a 99\% threshold, on average. If an order-of-magnitude improvement is made ($0.05\%$, green), these large-scale trees can be generated with a fidelity near 99.9\%.

\section{Extended Analysis of the Tree State Generator in a One-Way Quantum Repeater} \label{sec:repeater}
\begin{figure}[ht!]
\centering
\includegraphics{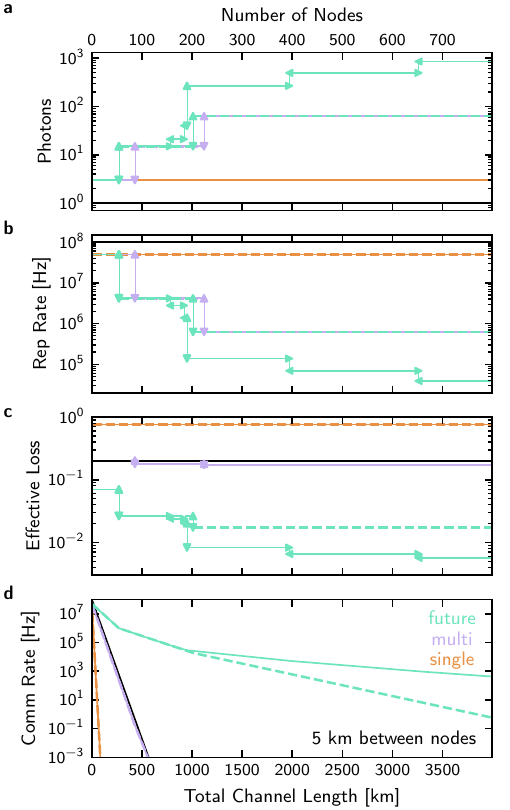}
\caption[Extended projections for a one-way quantum repeater based on a QPNN-based tree state generator]{\textbf{a} The number of photons in the optimal tree-type cluster state as a function of the total channel length (keeping a constant 5 km node separation; number of nodes shown on top axis) for the three different QPNN platforms (cf. Fig.~2). Dashed curves represent trees with a constant branching ratio $b=2$, while solid curves are arbitrarily-shaped trees with $\max\{\vec{b}\}\leq 4$. A change in the depth (branching) of the tree shape is denoted by vertical (horizontal) arrowheads. \textbf{b}-\textbf{d} The corresponding repetition rates, effective losses, and communication rates for the different clusters in \textbf{a}, benchmarked to the measures achieved using single photons (black lines). In this model, we assume that information is transferred between logical qubits perfectly at each node, such that all rate reduction is due to losses.}
\label{fig:boosting-full}
\end{figure}

Here, we include an alternate version of Fig.~5 that includes two additional panels, Fig.~\ref{fig:boosting-full}. These show the effective loss (cf. Eq.~7) and repetition rate ($1/\Delta t_T$, see Eq.~5) for the optimal trees at each channel length.
In this analysis, we continue to assume perfect source and detector efficiencies. At each node of the one-way quantum repeater, a QPNN-based tree state generator creates a new tree to meet the one incoming from the previous node. A Bell-state measurement is performed between these trees to transfer the logical qubit from the incoming tree to the new one, and we take this measurement to be perfect. It is worth noting that once a QPNN-based tree state generator is realized, it must also be feasible to realize a near-perfect QPNN-based Bell state analyzer \cite{Ewaniuk:23}. Altogether, for $N$ nodes the communication rate is simply calculated as $(1-\epsilon_\mathrm{eff})^N/\Delta t_T$.

\section{Three-Source Model of the Tree State Generator} \label{sec:three-source}
\begin{figure}[ht!]
\centering
\includegraphics{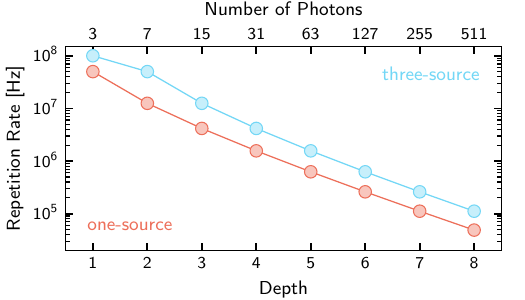}
\caption[Comparison between one-source and three-source versions of the QPNN-based generation protocol]{The repetition rate of the QPNN-based tree state generation protocol, when three (one) single-photon sources are used, is shown in blue (red) as a function of tree depth for 2-branch-only trees of increasing scale.}
\label{fig:three-source}
\end{figure}

As suggested in the main text, the repetition rate of the tree state generation protocol can be significantly increased if multiple single-photon sources are used, rather than just one. A simple example comes from using three single-photon sources when generating 2-branch-only trees, as were primarily discussed in the main text. If three sources are used, then the first stage of the regular protocol is effectively skipped. Instead of emitting each photon in the bottom of the tree sequentially, the protocol begins by emitting the bottom unit cell in its entirety such that the QPNN can begin entangling photons right away. Since the bottom row of any 2-branch-only tree contains more than half of its photons, this alteration boosts the repetition rate of the protocol by at least a factor of 2, as shown in Fig.~\ref{fig:three-source},
where this three-source version of the protocol is compared with the regular one-source version. In fact, for 7-photon and 15-photon trees, the repetition rate is increased by factors of 4 and 3, respectively. While the use of three sources in the initial stage of the protocol necessitates the use of additional switches to ensure the extra sources can be connected to the inputs of the QPNN, the delay line lengths and sizes of the other switches can be reduced. In sum, given that the overall loss is approximately the same, and the generation rate is directly proportional to the repetition rate, it is safe to say that this simple alteration would lead to about a factor of 2 rate enhancement in general.

\section{Extracting Directional Coupler Imbalance from State-of-the-Art Results} \label{sec:dc}
A MZI with a single phase shifter can be described by the following $2\times2$ transfer matrix,
\begin{equation} \label{eq:mzi}
    \mathbf{T}_\mathrm{mzi} = \begin{pmatrix}\sqrt{t_2}&-i\sqrt{1-t_2}\\-i\sqrt{1-t_2}&\sqrt{t_2}\end{pmatrix}\begin{pmatrix}e^{i2\theta}&0\\0&1\end{pmatrix}\begin{pmatrix}\sqrt{t_1}&i\sqrt{1-t_1}\\i\sqrt{1-t_1}&\sqrt{t_1}\end{pmatrix},
\end{equation}
where $2\theta$ is the phase shift and $t_1$, $t_2$ are the transmission coefficients of each directional coupler (ideally, $t_1=t_2=0.5$), respectively. When $\theta = 0$, the MZI will perfectly transmit photons passing through it such that those entering from the top arm will exit the bottom, and vice versa. Consider the imperfect directional coupler splitting ratio of 0.505 (i.e. 50.5\% transmission to 49.5\% reflection), and assume each is the same such that $t_1=t_2=0.505$. In this case, with $\theta=0$, the extinction ratio between reflection and transmission is calculated as 40 dB. It is evident from Fig.~2d in Ref.~\cite{Alexander:25} that the \textit{single}-platform has achieved more than 40 dB extinction for this $\theta$, and $>50$~dB for the opposite configuration (i.e. full reflection, $\theta=\frac{\pi}{2}$). Therefore, $0.5\%$ routing errors are the state-of-the-art (appropriate for \textit{multi} and \textit{future}) and have been demonstrated on the integrated platform we considered for our \textit{single} model.

\bibliography{references}